\documentclass[11pt,twoside]{article}

\begin{document}\hbadness=10000\thispagestyle{empty}
\title{Quantum fields, cosmological constant and symmetry doubling}


\author{Hans-Thomas Elze \\ \ \\ 
        Dipartimento di Fisica, Universit\`a di Pisa \\
        Largo Pontecorvo 3, I-56127 Pisa, Italia          \\ \ \\ 
        {\it E-mail: elze@df.unipi.it}}

\date{\today}

\maketitle

\begin{abstract}
We explore how energy-parity, a protective symmetry for the cosmological constant \cite{KS05}, arises naturally in the classical phase space dynamics of matter. We derive and generalize the Liouville operator of electrodynamics, incorporating a ``varying alpha'' and diffusion. In this model, a one-parameter deformation connects classical ensemble and quantum field theory. \\ 
{\it Keywords:} Emergent quantum theory, gauge symmetry, energy-parity. \\    
{\it PACS:} 03.65.Ta, 03.70+k, 05.20.-y
\end{abstract}

\section{Introduction}
The cosmological constant problem is one of the outstanding 
unsolved problems of physics \cite{Weinberg}. Besides approximate coincidence of its value  
with the average matter density of the universe in the present epoch, the 
problem consists in its smallness, $\Lambda\approx 10^{-123}M_{\rm Pl}^4$. One would expect    
corrections on the order of typical particle physics scales, for example, 
induced by interactions of the Standard Model. No cancellation 
mechanism with the required finesse is known.     

Addressing the smallness of $\Lambda$, the ``energy-parity'' symmetry introduced by 
Kaplan and Sundrum (KS) might protect the cosmological constant \cite{KS05}. 
Schematically, it concerns    
the mapping $Energy\rightarrow-Energy$, where the energy includes 
the contributions of all charged matter and gauge fields (summarily called  
matter henceforth). Correspondingly, the 
following effective low-energy Lagrangian coupling gravity and matter is considered: 
\begin{equation}\label{KSLagrangian} 
{\cal L}\equiv \sqrt{-g}\Big (M_{\rm Pl}^2R-\Lambda_0+{\cal L}_{mat}(\Phi )
-{\cal L}_{mat}(\tilde\Phi )\Big )
\;\;, \end{equation} 
where $g_{\mu\nu}$ denotes the metric entering the Einstein-Hilbert part of the 
Lagrangian, $\Lambda_0$ is the bare cosmological constant, and  
$\Phi$ stands for the set of all minimally coupled ``visible'' fields, 
while $\tilde\Phi$ denotes an identical ``ghost'' copy of the set of visible fields.  
The Lagrangian {\it function}, ${\cal L}_{mat}$, is the same for the visible and ghost sectors. 
 
The main point of Lagrangian (\ref{KSLagrangian}) is the relative sign between 
visible and ghost matter contributions. This leads to equal and opposite vacuum 
energies for both sectors. Therefore, they cancel and do not contribute to the bare 
and possibly small cosmological constant, $\Lambda_0$. 
Such a scenario has been 
earlier proposed by Linde \cite{Linde1} (in a modified form in Ref.\,\cite{Linde2}).   
Similar ideas based on symmetries have more recently  
also been discussed in 
Refs.\,\cite{Quiros04,Moffat05,Hossenfelder06,tHooftN06}.  
However, vacuum instabilities due to visible-ghost couplings threaten 
all models with fields that contribute negatively to the energy content of matter. 
KS have shown that vacuum decay may be 
acceptably slow under reasonable assumptions and is compatible with inflation 
and standard Big Bang cosmology \cite{KS05}.    

Our aim presently is threefold: 
\begin{itemize} 
\item 
To show that the {\it energy-parity symmetry} arises, if  
dynamics encoded in ${\cal L}_{mat}(\Phi )$ is described in phase space. 
For this purpose, we consider a Hilbert space representation of the  
Liouville equation. 
\item 
To show that this {\it classical Liouville equation}, written in terms of 
appropriate variables, is related 
to the {\it functional Schr\"odinger equation}, i.e. 
quantum field theory, pertaining to the matter Lagrangian, 
${\cal L}_{mat}(\Phi )-{\cal L}_{mat}(\tilde\Phi )$. However, we will also find 
here destabilizing visible-ghost couplings. 
\item 
To show that the visible-ghost couplings are eliminated by incorporating  
{\it diffusion} into the Liouville operator, 
consistently with energy-parity and gauge symmetry, and by introducing a varying gauge 
coupling, similarly as in ``varying alpha'' or dilaton models \cite{B82,Uzan03,KM04,Orfeu04}. 
This results in a one-parameter interpolation between the {\it classical ensemble theory} and 
the {\it quantum field theory} (QFT) for the matter Lagrangian 
in Eq.\,(\ref{KSLagrangian}). 
\end{itemize}
Thus, one might speculate about a relation between cosmological constant 
problem, energy-parity symmetry, ``varying alpha'' models, and recently studied 
{\it deterministic dynamics beneath quantum theory}. --  
The latter investigations have been initiated by work of 
't\,Hooft, motivated by 
the conceptual problems in unifying general relativity and 
quantum theory \cite{tHooft01,tHooft03,tHooft06}. 
 
There have always been arguments for and against the possibility  
to derive quantum theory from more fundamental and deterministic 
dynamical structures. The debate of hidden variables theories is well known. 
While much of this has come under experimental scrutiny, no deviation from quantum theory 
has been observed on the accessible scales.   
Nevertheless, it is quite plausible that quantum mechanics 
emerges as an effective theory only on sufficiently large scales compared to the Planck scale \cite{tHooft01}. 

Our approach makes use of the remarkable similarity between Schr\"odinger and 
Liouville equation, when written in terms of suitable variables  
\cite{I05susy}. The Liouville 
operator (times $i$) is Hermitian in the operator approach to classical statistical mechanics. 
One would like to identify it with the {\it Hamiltonian of an emergent quantum system}. 
However, unlike the case of a  
quantum mechanical Hamilton operator, 
its spectrum is generally {\it not bounded from below}.  
Therefore, attempts to find a deterministic foundation of quantum theory by relating it 
to a classical ensemble theory, so far, had  
to face the difficult problem of constructing a stable ground state 
\cite{tHooft01,tHooft03,I05susy,Vitiello01,ES02,I04,E04,Blasone04}. 

The simplest emergent quantum models are based on a classical system evolving  
in discrete time steps (cellular automaton) \cite{tHooft01,ES02}. 
It appears that all classical Hamiltonian models turn into unitary quantum mechanical 
ones, if the Liouville operator is discretized \cite{I04}. 
The arbitrariness inherent in discretizations leaves enough freedom for  
the construction of a ground state. However, interacting 
field theories, so far, have resisted to this.  


Various other arguments for deterministically induced quantum features have been presented -- 
see works collected in Part\,III of Ref.\,\cite{E04}, for example, or Refs. \cite{Smolin,Adler}, 
concerning statistical systems, quantum gravity, and matrix models. In detail, 
however, many of these incorporate variants of stochastic quantization procedures 
of Nelson and of 
Parisi and Wu based on an unknown mechanism driving the fluctuations \cite{Nelson}. 

Considering deterministic real-time evolution, 
we will show here that the characteristic doubling of classical 
phase space degrees of freedom, as compared to the quantum mechanical case, 
gives rise to the visible and ghost sectors of the KS energy-parity scenario. 
In this way, the previous difficulty stemming from  
the presence of the negative energy sector is turned into a virtue. However, 
destabilizing visible-ghost couplings arise and need to be rendered harmless.  

In Section\,2, we begin with a classical U(1) gauge  
theory, such as electrodynamics,  
with charged particles represented by complex Grassmann algebra valued fields 
\cite{I05susy}. We admit a variable  
gauge coupling as in the ``variable alpha'' or dilaton models \cite{B82,Uzan03,KM04}. 
From Hamilton's equations we obtain the Liouville equation. -- 
We develop the equivalent Hilbert 
space formulation, which automatically incorporates a 
ghost copy of the visible sector. 
This {\it classical} equation resembles a functional Schr\"odinger 
equation in which visible and ghost fields contribute with opposite sign to 
the emergent Hamiltonian. 

In Section\,3, we introduce a diffusion term   
which is compatible with gauge invariance and energy-parity symmetry.  
It uniquely incorporates only one extra derivative in the field variables  
without additional dimensionfull parameters. --   
Depending on the variable coupling, 
considered as a deformation parameter here, 
the {\it extended equation interpolates between a classical phase space ensemble theory and the functional Schr\"odinger equation}. In the latter 
limit, visible-ghost matter couplings are absent and 
KS energy-parity symmetry is manifest.       

In Section\,4, we briefly summarize and point out interesting 
topics for future study. 

\section{The Liouville operator equation for a classical gauge theory}
A charged matter field can be described by ``pseudoclassical mechanics''. 
This has been introduced through work of Casalbuoni and of Berezin and 
Marinov, who considered a {\it Grassmann variant of classical mechanics}, 
studying the dynamics of spin degrees of freedom classically and 
after quantization as usual \cite{CB}. For some recent applications, 
see Refs.\,\cite{I05susy,FDeW}, for example.   

Thus, we introduce the complex four-component spinor field, $\psi$, 
which takes its ``fermionic'' character from the generators of an 
infinite dimensional Grassmann algebra \cite{FDeW}. They obey:  
\begin{equation}\label{psi} 
\{ \psi (x),\psi (x')\}_+\equiv \psi (x)\psi (x')+\psi (x')\psi (x)=0 
\;\;, \end{equation}
where $x,x'$ are equal-time coordinate labels and spinor indices are suppressed. -- 
Furthermore, the usual four-vector potential, 
$A_\mu ,\;\mu =0,1,2,3$, defines the gauge field, 
$F_{\mu\nu}\equiv \partial_\mu A_\nu -\partial_\nu A_\mu$. --  
We assume the Minkowski metric $g_{\mu\nu}\equiv \mbox{diag}(+1,-1,-1,-1)$, since the 
essential aspects in the following do not depend on the background metric.  

Then, the {\it classical U(1) gauge theory} to be studied is defined by the action: 
\begin{eqnarray}\label{action}
S&\equiv&\int\mbox{d}^4x\;{\cal L}_{mat}(\psi ,A)
\\ [1ex] \label{action1}
&\equiv &\int\mbox{d}^4x\;\Big (\bar\psi (i\gamma\cdot D-m)\psi -\frac{1}{4\epsilon^2}F^2\Big )
\;\;, \end{eqnarray} 
where $\bar\psi_b \equiv \psi^*_{a}\gamma^0_{ab}$ ($\gamma$'s denoting the Dirac gamma matrices and  
$a,b=1,\dots,4$ spinor indices), 
$m$ is the mass parameter, and the covariant derivative is defined by $D_\mu \equiv \partial_\mu +ie_0A_\mu$. We  
also introduced here the scalar, dimensionless, and gauge neutral field $\epsilon$, such that the electric charge is given by $e(x)=e_0\epsilon (x^\mu )$ \cite{B82,KM04}. Thus, 
the field $\epsilon^{-1}$ plays the role of a dielectric function of the vacuum and, generally,   
varies in space and time. -- Models with varying coupling constants (or dilaton fields) 
have originated in various contexts \cite{B82,Uzan03,KM04,Orfeu04}. 
For our purposes, it is sufficient to consider $\epsilon$ as a variable 
{\it deformation parameter} of our example gauge theory.  
  
Proceeding as usual, we calculate the fermionic  
canonical momentum ($\int\mbox{d}^4x{\cal L}_{mat}\equiv\int\mbox{d}tL$): 
\begin{equation}\label{Ppsi}
\Pi \equiv -\frac{\delta L}{\delta\partial_0\psi }=i\psi^* 
\;\;, \end{equation} 
with a functional left derivative here. Concerning Grassmann variables, we will always use  
{\it left derivatives} \cite{FDeW}; for example, $\delta_\psi\equiv\delta_\psi^L$, 
with:
\begin{equation}\label{leftder} 
\delta_\psi^L(x)\psi (y)\psi (z)\equiv\delta^3(x-y)\psi (z)-\psi (y)\delta^3(x-z)
\;\;. \end{equation}
The momenta conjugate to the gauge potentials are:   
\begin{equation}\label{P0}
\Pi_0\equiv \frac{\delta L}{\delta\partial_0A^0}=0 
\;\;, \end{equation} 
\begin{equation}\label{Pi}
\Pi_i\equiv \frac{\delta L}{\delta\partial_0A^i}=-\epsilon^{-2}F_{0i}\equiv \epsilon^{-2}E^i 
\;\;. \end{equation} 
Then, we obtain the Hamiltonian:  
\begin{eqnarray}
H&=&\int\mbox{d}^3x\;\Big (\Pi\partial_0\psi +\Pi_i\partial_0A^i\Big )-L 
\nonumber \\ [1ex] \label{Hamiltonian}
&=&\int\mbox{d}^3x\;\Big (-\Pi\gamma^0(\gamma_jD^j+im)\psi +\frac{\epsilon^2}{2}\Pi_i\Pi_i
+\frac{1}{4\epsilon^2}F_{ij}F^{ij}
+A^0(\partial_i\Pi_i-ie_0\Pi\psi\Big )
\;\;, \end{eqnarray} 
after partially integrating the penultimate term, and summing over  
pairwise equal indices always. 

It is obvious from Eq.\,(\ref{P0}) and the term involving $A^0$ in Eq.\,(\ref{Hamiltonian}) 
that $A^0$ is the Lagrange multiplier which incorporates Gauss' law as a constraint: 
\begin{equation}\label{Gauss} 
G\equiv -\frac{\delta L}{\delta A^0}=\partial_i\Pi_i-ie_0\Pi\psi =0
\;\;. \end{equation} 
Keeping this constraint in mind, we work in temporal axial gauge,  
$A^0\equiv 0$, from now on. 

For the phase space description of the classical field theory, it will be useful to introduce the Poisson bracket operation, acting on two observables $O_1$ and $O_2$ which are function(al)s of the phase space variables 
$\Pi ,\psi ,\Pi_i ,A^i$ and may explicitly depend on time: 
\begin{equation}\label{PB} 
\{O_1,O_2\}\equiv \int\mbox{d}^3x\;\Big (     
\frac{\delta O_1}{\delta\Pi}\frac{\delta O_2}{\delta\psi}
+\frac{\delta O_1}{\delta\psi}\frac{\delta O_2}{\delta\Pi}
+\frac{\delta O_1}{\delta\Pi_i}\frac{\delta O_2}{\delta A^i}
-\frac{\delta O_1}{\delta A^i}\frac{\delta O_2}{\delta\Pi_i}
\Big )
\;\;, \end{equation} 
where all functional (left) derivatives refer to the same space-time argument. 
This Poisson bracket is {\it graded} antisymmetric: it is antisymmetric, 
if both observables $O_1,\,O_2$ are Grassmann even, it is symmetric, if 
both are odd, while in the remaining cases the terms involving Grassmann 
derivatives contribute symmetrically and the others antisymmetrically \cite{CB,FDeW}.

For any observable $O$, the usual relation among time derivatives holds: 
\begin{equation}\label{timederivs} 
\frac{\mbox{d}}{\mbox{d}x^0}O=\{ H,O\}+\partial_0O
\;\;, \end{equation}
which embodies Hamilton's equations of motion. For example, $\mbox{d}\psi /\mbox{d}x^0=\{H,\psi\}$ and 
$\mbox{d}\Pi /\mbox{d}x^0=\{H,\Pi\}$, i.e. the Dirac equation and its adjoint.  
The time independent Hamiltonian of Eq.\,(\ref{Hamiltonian}) is conserved. 
One also verifies that $\{ H,G\}=0$,   
expressing the gauge invariance of the evolution of the system. Consequently,    
it is sufficient to implement Gauss' law, Eq.\,(\ref{Gauss}), at one time.     
-- We now turn to the study of an ensemble of systems 
that are described by Eq.\,(\ref{action1}). 

\subsection{The Hilbert space representation}  
A particular example of Eq.\,(\ref{timederivs}) is the Liouville equation 
for a conservative system.   
Considering an ensemble, especially with some distribution over initial conditions, 
this equation governs the evolution of its phase space density $\rho$: 
\begin{equation}\label{Liouville} 
0=\frac{\mbox{d}}{\mbox{d}x^0}\rho =\partial_0\rho -\hat{\cal L}\rho
\;\;, \end{equation}
\begin{equation}\label{Liouvilleop}  
-\hat{\cal L}\rho \equiv \{ H,\rho \}
\;\;, \end{equation}
where $\hat{\cal L}$ is the Liouville operator.  
These equations summarize the classical statistical mechanics of a conservative system, 
given the Hamiltonian $H$.  

An equivalent Hilbert space formulation is obtained in the operator approach,  
with appropriate 
modifications for our classical field theory, which is based on two ingredients: 
\begin{enumerate} 
\item The phase space density functional can be factorized in the form $\rho\equiv\Psi^*\Psi$. 
This surprising fact is guaranteed by a theorem proven by 't\,Hooft recently 
\cite{tHooft06}. 
\item The Grassmann algebra valued complex state functional $\Psi$ itself obeys 
Eq.\,(\ref{Liouville}). 
\end{enumerate} 
Furthermore, the complex valued inner product of such state functionals is defined by: 
\begin{equation}\label{scalarprod}
\langle\Psi |\Phi\rangle \equiv \int{\cal D}\Pi{\cal D}\psi{\cal D}\Pi_j{\cal D}A^i\; 
\Psi^*\Phi =\langle\Phi |\Psi\rangle^*
\;\;, \end{equation}
functionally integrating over all phase space variables (fields). 
-- Due to the presence of Grassmann variables, the $*$-operation which defines  
the dual of a state functional needs special attention.  
It amounts to complex conjugation for a ``bosonic'' 
state functional, $(\Psi [\Pi_j,A^i])^*\equiv \Psi^*[\Pi_j,A^i]$, analogously to    
an ordinary wave function in quantum mechanics. 

However, based on complex conjugation alone, 
the inner product involving Grassmann variables would not be well defined, in particular, it 
would not necessarily yield a complex number nor a positive definite norm.
Instead, a detailed construction of the inner 
product for functionals of Grassmann algebra valued fields has been carried out in 
Ref.\,\cite{Jackiw}, which has these physically motivated properties. 
Other constructions are possible \cite{BarnesGhandour}; 
see also the discussion in the Appendix of Ref.\,\cite{Jackiw}. Further applications 
can be found in Refs.\,\cite{Kiefer}.  

Thus, the notion of square-integrable functions can be extended to 
the phase space functionals (rigorously after discretization).
Furthermore, only ``physical'' state functionals which conform with Gauss' law 
(cf. below) are admitted here. 

Given the Hilbert space structure, the operator $\hat{\cal H}\equiv i\hat{\cal L}$ 
has to be Hermitian for a conservative system 
and the positive overlap $\langle\Psi |\Psi\rangle$ is a conserved quantity. 
Then, the Liouville equation also applies to 
$\rho =\Psi^*\Psi$, due to its linearity. Interpreting $\rho$ as the phase space distribution, 
its moments yield the physically meaningful expectation values of observables, 
as usual. 

The similarity with the usual quantum mechanical formalism is striking. In order to 
expose this more clearly, we further transform  
the functional equation implied by the above two postulates together with 
Eqs.\,(\ref{Liouville})--(\ref{Liouvilleop}).    
Let us write this equation in the suggestive form: 
\begin{equation}\label{Schroedinger} 
i\partial_t\Psi =\hat{\cal H}\Psi
\;\;, \end{equation} 
where $\Psi$ is a functional of $\Pi,\psi,\Pi_i,A^i$, and where  
the effective ``Hamilton operator'' is: 
\begin{equation}\label{Hem} 
\hat{\cal H}\Psi =-i\{ H,\Psi\}\equiv (\hat{\cal H}_\psi +\hat{\cal H}_A)\Psi
\;\;. \end{equation}
The operators $\hat{\cal H}_\psi$ and $\hat{\cal H}_A$, respectively, refer to terms 
which originate from the Poisson bracket either involving Grassmann derivatives or not, 
see Eq.\,(\ref{PB}). We consider both terms in turn. 

\subsection{The gauge field operator $\hat{\cal H}_A$} 
Beginning with the gauge field part, we obtain: 
\begin{equation}\label{HemA}
\hat{\cal H}_A\Psi
=-i\int\mbox{d}^3x\;\Big (\epsilon^2\Pi_i\frac{\delta}{\delta A^i}
-[\frac{1}{\epsilon^2}(\partial^jF_{ij})-ie_0\Pi\gamma^0\gamma_i\psi ]
\frac{\delta}{\delta\Pi_i}\Big )\Psi 
\;\;. \end{equation}
The form of the kinetic term suggests to perform a functional Fourier transformation: 
\begin{equation}\label{Fourier}
\Psi [\Pi_i]=\int {\cal D}A'^i\;\exp (-i\Pi_i\cdot A'^i)\Psi [A'^i]
\;\;, \end{equation} 
where all other variables are momentarily suppressed; in the 
exponent, an integration over space is understood. 
In the transformed variables, the Eq.\,(\ref{Hem}) reads: 
\begin{equation}\label{HemA1}
\hat{\cal H}_A\Psi
=\int\mbox{d}^3x\;\Big (-\epsilon^2\frac{\delta}{\delta A^i}\frac{\delta}{\delta A'^i}
+\frac{1}{2\epsilon^2}F_{ij}F'^{ij}   
-ie_0\Pi\gamma^0\gamma_i\psi A'^i\Big )\Psi
\;\;, \end{equation}
making use of suitable partial integrations, and 
where $F'^{ij}\equiv \partial^iA'^j-\partial^jA'^i$. Next, 
we perform a linear transformation of the gauge field variables: 
\begin{equation}\label{transform} 
A^i\equiv \frac{1}{\sqrt 2}(a^i+\tilde a^i)\;\;,\;\;\;
A'^i\equiv \frac{1}{\sqrt 2}(a^i-\tilde a^i)
\;\;. \end{equation}
This transforms the operator of Eq.\,(\ref{HemA1}) into: 
\begin{equation}\label{HemA2} 
\hat{\cal H}_A
=\int\mbox{d}^3x\;\Big (-\frac{\epsilon^2}{2}(
\frac{\delta}{\delta a^i}\frac{\delta}{\delta a^i}
-\frac{\delta}{\delta\tilde a^i}\frac{\delta}{\delta\tilde a^i})
+\frac{1}{4\epsilon^2}(F_{ij}F^{ij}-\tilde F_{ij}\tilde F^{ij})   
-ie'_0\Pi\gamma^0\gamma_i\psi (a^i-\tilde a^i)\Big ) 
\;\;, \end{equation}
where $e'_0\equiv e_0/\sqrt 2$, $F^{ij}\equiv \partial^ia^j-\partial^ja^i$, 
and $\tilde F^{ij}\equiv \partial^i\tilde a^j-\partial^j\tilde a^i$. -- 
Thus, we find that a ghost copy in terms of $\tilde a$ arises here with opposite 
sign for each visible gauge field term involving $a$.   

\subsection{The fermionic field operator $\hat{\cal H}_\psi$}
Proceeding with the fermionic field part, we obtain: 
\begin{equation}\label{Hempsi}
\hat{\cal H}_\psi\Psi  
=\int\mbox{d}^3x\;\Big (
-\psi\big [\gamma^0(-i\gamma_j\stackrel{\leftharpoonup}{D'^j}+m)\big ]^t\delta_\psi 
+\Pi\gamma^0(-i\gamma_j\stackrel{\rightharpoonup}{D'^j}+m)\delta_\Pi
\Big )\Psi  
\;\;, \end{equation} 
indicating which way the derivatives act, 
with $D'^j\equiv \partial^j +ie'_0(a^j+\tilde a^j)$, and where $\big [\dots\big ]^t$ 
denotes spinor matrix transposition. Explicitly, the   
first term is:  
$\psi M^t\delta_\psi\equiv \psi_b(M^t)_{ba}\delta_{\psi_a}=M_{ab}\psi_b\delta_{\psi_a}$. 

Making use of 
the algebra of $\gamma$-matrices -- in particular, the  
{\it charge conjugation} matrix $C$ 
and the matrix $\gamma_5$, with $C\gamma_\mu C^{-1}=-\big [\gamma_\mu\big ]^t$, $C^2=-I$,  
$\{\gamma_5,\gamma_\mu\}_+=0$, and $\gamma_5^{\;2}=I$ \cite{IZ} --   
and of a partial integration in the first term, we rewrite the   
operator $\hat{\cal H}_\psi$ as: 
\begin{equation}\label{Hempsi1} 
\hat{\cal H}_\psi  
=\int\mbox{d}^3x\;\big (
-\psi_Ch_D^-\delta_{\psi_C}+\Pi h_D^+\delta_\Pi\big )  
\;\;, \end{equation}
with: 
\begin{eqnarray}\label{psiC} 
\psi_C&\equiv&\psi (\gamma_5C)^{-1} 
\;\;, \\ [1ex]\label{hD} 
h_D&\equiv&\gamma^0(-i\gamma_jD'^j+m)
\;\;, \end{eqnarray} 
i.e., the Hermitian kernel of the Dirac Hamiltonian,  
with the superscript ``$\pm$'' indicating the sign of the minimal coupling term in  
the covariant derivative. 

We may go one step further, employing the {\it time reversal} matrix $T$ \cite{IZ}. It is characterized by 
the relation $T\gamma_\mu T^{-1}=\gamma^\mu$ and involves complex conjugation, such that 
$T(i\gamma_\mu )T^{-1}=-i\gamma^\mu$, for example. Thus, the sign of the coupling in the 
first term in Eq.\,(\ref{Hempsi1}) can be changed: 
$Th_D^-T^{-1}=h_D^+$. 
Correspondingly, we introduce: 
\begin{equation}\label{psiCT} 
\psi_{CT}\equiv\psi_CT^{-1}=\psi (T\gamma_5C)^{-1}  
\;\;, \end{equation}  
and obtain: 
\begin{equation}\label{Hempsi2} 
\hat{\cal H}_\psi =\int\mbox{d}^3x\;
\big (-\psi_{CT}h_D\delta_{\psi_{CT}}+\Pi h_D\delta_\Pi\big )  
\;\;, \end{equation}
with $h_D=h_D^+$, from now on. 

As suggested by Eqs.\,(\ref{HemA2}) and (\ref{Hempsi2}), we    
interpret $\Pi$ as ghost copy of $\psi_{CT}$.  
Summarizing our findings, we state the 
{\it energy-parity symmetry} transformations: 
\begin{eqnarray}\label{ep1}
\psi_{CT}&\longleftrightarrow&\Pi
\;\;, \\ [1ex] \label{ep2} 
a_i&\longleftrightarrow&\tilde a_i 
\;\;. \end{eqnarray} 
Applying these to the effective Hamiltonian operator, 
$\hat{\cal H}=\hat{\cal H}_\psi +\hat{\cal H}_A$, we obtain indeed: 
\begin{equation}\label{ep3}
\hat{\cal H}\;\longleftrightarrow\;-\hat{\cal H} 
\;\;. \end{equation} 
A little algebra is needed, in order to show that 
the interaction term $\propto e'_0$ in $\hat{\cal H}_A$, see Eq.\,(\ref{HemA2}), 
conforms with (\ref{ep3}); in particular, $\big [T\gamma_5C\big ]^t=-T\gamma_5C$. 

This result demonstrates that the energy-parity symmetry  
which has been postulated 
by Kaplan and Sundrum \cite{KS05} arises naturally 
in the Hilbert space representation of the classical ensemble theory 
for our U(1) model of Eq.\,(\ref{action}). 

\subsection{Gauge invariance}  
Let us briefly discuss the operator version of 
Gauss' law, Eq.\,(\ref{Gauss}).  
Similarly as the effective Hamiltonian in  
the previous subsections, we obtain the {\it Gauss' law operator}: 
\begin{eqnarray}\label{Gaussop} 
\hat{\cal G}\Psi&\equiv &-i\{G,\Psi\}/\sqrt 2
\\ [1ex] \label{Gaussopdet}
&=&\big (\frac{i}{2}\partial_i(\delta_{a^i}+\delta_{\tilde a^i})
-e'_0\psi\delta_\psi +e'_0\Pi\delta_\Pi\big )\Psi  
\;\;, \end{eqnarray} 
with an extra factor of $\sqrt 2$ for later convenience.
Locally, we then have the Gauss' law constraint: 
\begin{equation}\label{Gaussopconstr}
\hat{\cal G}(x)\Psi =0 
\;\;, \end{equation} 
which needs to be implemented together with the implicit gauge 
condition $A^0=0$, in order to eliminate unphysical gauge degrees of freedom. --  
Finally, a straightforward calculation yields the commutator: 
\begin{equation}\label{gaugeinv}
[\hat{\cal H},\hat{\cal G}(x)]\Psi=0
\;\;, \end{equation}
which, again, expresses the local U(1) gauge invariance of the evolution 
generated by $\hat{\cal H}$. Therefore, the constraint can be implemented consistently 
as an initial condition, for example. 
  
This completes the derivation of classical statistical mechanics for a U(1) 
gauge field theory in a Hilbert space formalism, uncovering the intrinsic energy-parity 
symmetry.

\section{``Varying alpha'', diffusion and transition to QFT}
It will be shown here that the classical ensemble theory of Section\,2 presents 
a limit of a more general ensemble theory. In another limit, varying the deformation 
parameter $\epsilon$, this reduces to a quantized gauge field theory    
in the Schr\"odinger picture.  

We reparametrize $\epsilon$, as 
introduced in the action, Eq.\,(\ref{action}), for      
variable coupling \cite{B82,Uzan03}: 
\begin{equation}\label{epsprime} 
\epsilon^2\equiv\epsilon^2(\epsilon')\equiv \frac{1}{2}\epsilon'(1+\epsilon')
\;\;. \end{equation} 
Furthermore, the following linear transformation of the vector fields is implemented: 
\begin{eqnarray}\label{aplus}
a+\tilde a&\equiv &\frac{1}{2}(1+\epsilon')(a'+\tilde a')
\;\;, \\ [1ex] \label{aminus} 
a-\tilde a&\equiv &\epsilon'(a'-\tilde a')
\;\;. \end{eqnarray}
Contributions to the effective Hamilton operator, 
$\hat{\cal H}=\hat{\cal H}_\psi +\hat{\cal H}_A$, previously obtained in 
Eqs.\,(\ref{HemA2}) and (\ref{Hempsi}), respectively, are transformed 
accordingly. We obtain:
\begin{eqnarray}
\hat{\cal H}_A
&=&\int\mbox{d}^3x\;\Big (-\frac{1}{2}(
\frac{\delta}{\delta a'^i}\frac{\delta}{\delta a'^i}
-\frac{\delta}{\delta\tilde a'^i}\frac{\delta}{\delta\tilde a'^i})
+\frac{1}{4}\big (F_{ij}(\epsilon')F^{ij}(1+\epsilon')
-\tilde F_{ij}(\epsilon')\tilde F^{ij}(1+\epsilon')\big )
\nonumber \\ [1ex] \label{HemA3} 
&\;&\;\;\;\;+\frac{1}{4(1+\epsilon')}\big (F_{ij}(\epsilon')\tilde F^{ij}(1)
-F_{ij}(1)\tilde F^{ij}(\epsilon')\big )
-ie'_0\epsilon'\Pi\gamma^0\gamma^i\psi (a'_i-\tilde a'_i)\Big )
\;\;, \end{eqnarray}
with $F^{ij}(\xi )\equiv\xi^{-1}(\partial^i\xi a^j-\partial^j\xi a^i)$, 
and $\tilde F^{ij}(\xi )\equiv\xi^{-1}(\partial^i\xi \tilde a^j-\partial^j\xi \tilde a^i)$; 
the latter present the appropriate generalization of the usual 
field strength tensor \cite{B82,Uzan03,KM04}, 
which is recovered in the case of spatially constant $\xi$. --  
Finally, the covariant derivative transforms to:  
\begin{equation}\label{covderiv}
D'^j\equiv \partial^j +i\frac{e'_0}{2}(1+\epsilon')(a'^j+\tilde a'^j) 
\;\;, \end{equation}
and, with this, the fermionic contribution, $\hat{\cal H}_\psi$, 
{\it retains its form}, as before in Eqs.\,(\ref{Hempsi}) or (\ref{Hempsi2}). 

The varying {\it effective charge} $e_{eff}\equiv e'_0(1+\epsilon')/2$, which is seen 
in Eq.\,(\ref{covderiv}), also enters the correspondingly transformed 
Gauss' law operator from Eq.\,(\ref{Gaussopdet}): 
\begin{equation}\label{Gaussop1} 
(e'_0)^{-1}\hat{\cal G}
=\frac{i}{2}\partial_ie_{eff}^{-1}(\delta_{a'^i}+\delta_{\tilde a'^i})
+\psi_{CT}\delta_{\psi_{CT}} +\Pi\delta_\Pi  
\;\;. \end{equation}   
Note that $\psi i\delta_\psi =-\psi_{CT}\delta_{\psi_{CT}}$, 
cf. Eqs.\,(\ref{Gauss}), (\ref{psiCT}). -- So far, this is still 
the Hilbert space version of statistical mechanics 
and Liouville equation, in particular, for our classical field theory.  

\subsection{Generalization of the Liouville equation}
We now incorporate an {\it additional interaction} term, $\hat{\cal H}_{int}$, into   
the effective Hamilton operator:  
\begin{equation}\label{Hfinal} 
\hat{\cal H}\equiv\hat{\cal H}_\psi +\hat{\cal H}_A+\hat{\cal H}_{int}
\;\;, \end{equation} 
\begin{equation}\label{Hint}  
\hat{\cal H}_{int}  
\equiv -\int\mbox{d}^3x\;\Big (\frac{e'_0}{2}(1-\epsilon')(a'^j-\tilde a'^j)
\big (
\psi\big [\gamma^0\gamma_j\big ]^t\delta_\psi
+\Pi\gamma^0\gamma_j\delta_\Pi\big )\Big )  
\;\;. \end{equation} 
This should be compared to the minimal coupling terms in $\hat{\cal H}_\psi$, 
Eq.\,(\ref{Hempsi}), 
inserting the covariant derivative of Eq.\,(\ref{covderiv}); using  
$\psi\big [\gamma^0\gamma_j\big ]^t\delta_\psi =\psi_{CT}\gamma^0\gamma_j\delta_{\psi_{CT}}$, 
yields the appropriate term for comparison with Eq.\,(\ref{Hempsi2}). -- 
Since $\hat{\cal H}\equiv i\hat{\cal L}$, 
this generalizes the classical ensemble theory.    

Several remarks are in order here characterizing the new term $\hat{\cal H}_{int}$:
\begin{itemize} 
\item It vanishes in the limit $\epsilon'\rightarrow 1$, where Eq.\,(\ref{Hfinal}) 
presents the Hilbert space 
version of the Liouville operator (times $i$) of the classical theory with 
gauge coupling constant $e_0$.
\item It involves one extra phase space derivative as compared to  
all other terms in $\hat{\cal H}$, which were generated by the Poisson bracket, Eq.\,(\ref{PB}), 
without introducing an additional dimensionfull parameter.    
Recall that $a'^j-\tilde a'^j\propto i\delta /\delta\Pi_j$, by Fourier transformation, 
where $\Pi_j$ is the original canonical momentum 
variable conjugate to $A^j$.
\item Since it is Hermitian (cf. below), it follows that $\hat{\cal H}_{int}$ 
represents a {\it diffusive interaction}.
\item It is gauge invariant, since $[\hat{\cal H}_{int},\hat{\cal G}]=0$, and conforms 
with energy-parity, Eqs.\,(\ref{ep1})--(\ref{ep3}).  
\end{itemize}

\subsection{Unitarity}
In order that the time evolution of the system, described by the functional $\Psi$, be unitary,  
the Hamilton operator $\hat{\cal H}$ needs to be Hermitian. 
 
The pure gauge field part of $\hat{\cal H}_A$,   
Eq.\,(\ref{HemA3}), fullfills this. It represents the Hamilton operator 
of two interacting {\it quantized gauge fields}, $a_i$ and ghost copy 
$\tilde a_i$, in the Schr\"odinger picture. Their interaction  
is solely due to the {\it spatial variation} of the coupling, since the 
terms $\propto (1+\epsilon')^{-1}$ in Eq.\,(\ref{HemA3}) cancel for constant $\epsilon'$. -- 
The construction of the corresponding function space, on which these bosonic operators act, 
follows the scalar field case reviewed in Ref.\,\cite{Jackiw}. 

Turning to the Grassmann variables, we now consider the field $\psi_{CT}$ and the 
functional derivative $\delta_{\psi_{CT}}$ as a 
representation of a {\it charged fermion field operator} $\hat\psi$ 
and its adjoint $\hat\psi^\dagger$, respectively. This is suggested by  
the equal-time anticommutation relations: 
\begin{equation}\label{fermionop}
\big\{\psi (x),\frac{\delta}{\delta\psi (x')}\big\}_+
=\delta^3(x-x')
=\big\{\hat\psi (x),\hat\psi^\dagger (x')\big\}_+
\;\;, \end{equation}
suppressing spinor indices, and  
analogously for $\Pi,\delta_\Pi$ and $\hat\Pi,\hat\Pi^\dagger$. Symbolically, we relate:  
\begin{eqnarray}\label{psiassoc}
(\psi_{CT};\delta_{\psi_{CT}})&\longleftrightarrow&(\hat\psi;\hat\psi^\dagger ) 
\;\;, \\ [1ex]\label{piassoc}   
(\Pi;\delta_\Pi)&\longleftrightarrow&(\hat\Pi;\hat\Pi^\dagger )
\;\;. \end{eqnarray} 
Thus, the subscript ``$\dots_{CT}$'' is absorbed in the overhead ``$\hat{\dots}$'' in what follows.  

Let us specify in more detail the space of functionals, on which 
the fermionic field operators act. Decomposing the fields $\psi,\Pi$  
into real and imaginary parts: 
\begin{equation}\label{psipisplit}
\psi\equiv\frac{1}{\sqrt 2}(\psi_R+i\psi_I)\;\;,\;\;\;\Pi\equiv\frac{1}{\sqrt 2}(\Pi_R+i\Pi_I)
\;\;, \end{equation} 
we associate operators with the real components: 
\begin{eqnarray}\label{psipiR}
\hat\psi_R&\equiv&\frac{1}{\sqrt 2}(u+\delta_u)\;\;,\;\;\;
\hat\Pi_R\;\equiv\;\frac{1}{\sqrt 2}(\tilde u+\delta_{\tilde u})  
\;\;, \\ [1ex]\label{psipiI}
\hat\psi_I&\equiv&\frac{1}{i\sqrt 2}(u-\delta_u)\;\;,\;\;\;
\hat\Pi_I\;\equiv\;\frac{1}{i\sqrt 2}(\tilde u-\delta_{\tilde u})
\;\;, \end{eqnarray} 
where $u,\tilde u$ are real (four-component) Grassmann fields.  

These operators act on functionals $\Psi [u,\tilde u]$. Constructing the dual $\Psi^*[u,\tilde u]$ as in 
Refs.\,\cite{Jackiw,BarnesGhandour}, 
the adjoints of $u$ and $\tilde u$ are $\delta_u$ and $\delta_{\tilde u}$, respectively.  
Therefore, the operators $\hat\psi_{R,I},\hat\Pi_{R,I}$ are Hermitian here.  
It follows that $\hat\psi =u$ and $\hat\psi^\dagger=\delta_u$, 
as well as $\hat\Pi =\tilde u$ and $\hat\Pi^\dagger=\delta_{\tilde u}$. In this way, the anticommutator 
relations (\ref{fermionop}) are realized, and analogous ones for $\hat\Pi,\hat\Pi^\dagger$. 

We adopt the convention of Berezin and Marinov that the {\it adjoint of a product}  
of Grassmann variables $\xi_1,\xi_2$, incorporating complex conjugation,  
is: $(\xi_1\xi_2)^\dagger =\xi_2^\dagger\xi_1^\dagger$ \cite{CB}. With this, the   
space of functionals, and the above realizations of the fields as operators, we have: 
\begin{eqnarray}\label{herm1} 
(\hat{\cal H}_\psi +\hat{\cal H}_{int})^\dagger 
&=&\hat{\cal H}_\psi +\hat{\cal H}_{int} 
\;\;, \\ [1ex]\label{herm2} 
\hat{\cal G}^\dagger 
&=&\hat{\cal G} 
\;\;. \end{eqnarray}
see Eqs.\,(\ref{Hempsi2}), (\ref{Hint}), and (\ref{Gaussop1}).  
Of course, the functionals $\Psi [u,a_i;\tilde u,\tilde a_i]$, generally, depend on all 
bosonic and fermionic fields, interpreted as visible ($u,a_i$) and ghost matter ($\tilde u,\tilde a_i$).  

One term of $\hat{\cal H}$ is left to be considered. 
Coupling fermions and bosons, see $\hat{\cal H}_A$ in Eq.\,(\ref{HemA3}): 
\begin{equation}\label{HApsipi} 
\hat{\cal H}_{A\psi\Pi}\equiv -ie'_0\epsilon'\Pi\gamma^0\gamma^i\psi (a'_i-\tilde a'_i)
\;\;, \end{equation}  
this, in general, is {\it not} Hermitian. Instead, the relevant factor of this term gives:   
\begin{equation}\label{herm3} 
(i\hat\Pi\gamma^0\gamma^iT\gamma_5C\hat\psi)^\dagger 
=-i(T\gamma_5C\hat\psi)^\dagger\gamma^0\gamma^i\hat\Pi^\dagger
\neq i\hat\Pi\gamma^0\gamma^iT\gamma_5C\hat\psi
\;\;, \end{equation} 
where we used Eq.\,(\ref{psiCT}) and (\ref{psiassoc}),\,(\ref{piassoc}). 

There are several ways to handle this situation, with different physical 
implications. -- 
First, by adding $\hat{\cal H}_{A\psi\Pi}^\dagger$ to $\hat{\cal H}$, we can 
make the resulting Hamilton operator Hermitian, maintaining 
gauge invariance and energy-parity symmetry. 
In terms of the original phase 
space variables, this term incorporates three functional derivatives. 
It corresponds to a generalization 
of the Liouville operator, similarly as adding $\hat{\cal H}_{int}$,  
Eq.\,(\ref{Hint}). While the latter is necessary for a smooth 
transition to quantum theory, as we shall see, we have no particular reason 
for the former addition. -- 
Second, we can impose a constraint: 
\begin{equation}\label{constr1} 
\hat{\cal C}_1\Psi\equiv\big (\hat\Pi -i(T\gamma_5C\hat\psi)^\dagger\big )\Psi =0 
\;\;, \end{equation}  
or a less restrictive constraint: 
\begin{equation}\label{constr2} 
\hat{\cal C}_2\Psi\equiv\big (\hat{\cal H}_{A\psi\Pi}-\hat{\cal H}_{A\psi\Pi}^\dagger\big )\Psi =0 
\;\;, \end{equation} 
and thereby reduce the Hilbert space, eliminating states that give rise to 
the anti-Hermitian part of $\hat{\cal H}_{A\psi\Pi}$. -- 
The constraint $\hat{\cal C}_1$ reminds us of the classical Eq.\,(\ref{Ppsi}). It eliminates 
ghost fermions as an independent field. In fact, there is a realization of the Hilbert space operators, 
differing from Eqs.\,(\ref{psiassoc})--(\ref{psipiI}), which automatically incorporates this constraint. 
Since it correlates ghost with visible 
fermions, even in the absence of interactions, it is not usefull here. 
The constraint $\hat{\cal C}_2$ eliminates only states that allow 
certain transitions between visible and ghost matter, in an ad hoc way. 
Since $[\hat{\cal H},\hat{\cal C}_{1,2}]\neq 0$, neither constraint can be simply 
imposed on initial conditions. -- 
Third, much less restrictive is the constraint: 
\begin{equation}\label{Cexpect} 
\langle\Psi |\hat{\cal C}_{1,2}|\Psi\rangle =0
\;\;, \end{equation} 
which allows fluctuations away from the strict constraints, 
Eqs.\,(\ref{constr1}), (\ref{constr1}). 
In this case, the evolution of the system is not unitary. In particular, 
in the limit $\epsilon'\rightarrow 1$, which concerns the classical ensemble theory, 
the larger Hilbert space here contains states which experience dissipative forces. 
This is interesting in view of the ``information loss'' ideas of  
Refs.\,\cite{tHooft01,Vitiello01,Blasone04}, which we mentioned in Section\,1, and we will 
come back to it.  
    
Finally, however, we observe that the non-Hermitian term $\hat{\cal H}_{A\psi\Pi}$ vanishes 
in the limit $\epsilon'\rightarrow 0$. This limit, in particular, will 
be further studied in what follows.  
   
\subsection{Symmetry doubling, energy-parity and quantum fields}
Considering the {\it varying gauge coupling}  
induced by $\epsilon$, as introduced in Eq.\,(\ref{action1}) 
and reparametrized in terms of  
$\epsilon'$ in Eq.\,(\ref{epsprime}), the underlying mechanism 
is {\it not} our present concern. It should be   
a low-energy reflection of a more fundamental theory than 
described by Eq.\,(\ref{KSLagrangian}) and is amply discussed 
in the literature \cite{B82,Uzan03,KM04,Orfeu04}. 
  
However, the reader not wishing to adopt such an idea   
is invited to take the following as remarks on a {\it one-parameter 
deformation} of the classical ensemble or quantum field theory, respectively, 
for our U(1) model. This possibility in itself seems interesting.  
  
As we have discussed already the ``pseudoclassical'' limit 
$\epsilon'\rightarrow 1$, we now study in more detail the limit 
$\epsilon'\rightarrow 0$. In this case, assuming that $\epsilon'$ is 
spatially homogeneous and collecting terms from Eqs.\,(\ref{Hempsi2}), 
(\ref{HemA3}), (\ref{covderiv}), (\ref{Hint}), 
according to Eq.\,(\ref{Hfinal}) and (\ref{psiassoc}), (\ref{piassoc}), 
we obtain: 
\begin{equation}\label{Hamop} 
\hat{\cal H}
=\int\mbox{d}^3x\;\Big (\hat\psi^\dagger H_D\hat\psi
-\frac{1}{2}\frac{\delta}{\delta a'^i}\frac{\delta}{\delta a'^i}
+\frac{1}{4}F_{ij}F^{ij}
-\big (\hat\Pi^\dagger \tilde H_D\hat\Pi
-\frac{1}{2}\frac{\delta}{\delta\tilde a'^i}\frac{\delta}{\delta\tilde a'^i}
+\frac{1}{4}\tilde F_{ij}\tilde F^{ij}\big )\Big )
, \end{equation}
with the kernel of the Dirac Hamiltonian: 
\begin{equation}\label{HD}
H_D\equiv\gamma^0\big (-i\gamma_j(\partial^j+ie'_0a^j)+m\big )
\;\;, \end{equation}
and where $\tilde H_D$ has $a^j$ replaced by $\tilde a^j$. 
This is the {\it Hamilton operator} of an U(1) gauge theory (fields $\psi,a_i$) in 
the Schr\"odinger picture, 
in temporal axial gauge, together with an identical 
ghost copy (fields $\Pi,\tilde a_i$). -- 
Thus, we have derived the quantized matter part of the KS model  
of Eq.\,(\ref{KSLagrangian}) for Abelian gauge symmetry \cite{KS05} .   

While the usual interaction remains (coupling $e'_0$),  
destabilizing visible-ghost matter couplings are absent in this ``quantum limit''.   
Of course, it is related to the particular additional interaction, $\hat{\cal H}_{int}$, 
introduced in Eq.\,(\ref{Hint}). Further, note that the  
Gauss' law operator of Eq.\,(\ref{Gaussop1}) decomposes into visible and ghost matter parts: 
\begin{equation}\label{Gaussop2} 
\hat{\cal G}
=i\partial_i\delta_{a'^i}+e'_0\hat\psi\hat\psi^\dagger 
+i\partial_i\delta_{\tilde a'^i}+e'_0\hat\Pi\hat\Pi^\dagger  
\equiv \hat{\cal G}_{vis}+\hat{\cal G}_{gho} 
\;\;, \end{equation}   
which, in the present limit, obey: 
\begin{equation}\label{doublsymm}
[\hat{\cal H},\hat{\cal G}_{vis}]=[\hat{\cal H},\hat{\cal G}_{gho}]=0 
\;\;. \end{equation}  
Thus, we find here the {\it doubled symmetry} U(1$)_{vis}$\,x\,U(1$)_{gho}$, 
in agreement with the KS scenario. 

Furthermore, 
the quantum limit has the following features seen in Eq.\,(\ref{action1}) or 
Eqs.\,(\ref{epsprime})--(\ref{aminus}), where correspondingly 
$\epsilon'\rightarrow 0$ and $\epsilon^2(\epsilon')\rightarrow 0$. Undoing the linear 
and Fourier transformations involved, the gauge field (operators) are related to    
the original phase space variables: 
\begin{eqnarray}\label{correspondence1} 
a'_j&\sim&\frac{i}{\epsilon'(\epsilon )}\delta_{\Pi^j}+\frac{1}{1+\epsilon'(\epsilon )}A_j
\\ [1ex] \label{correspondence2}
\tilde a'_j&\sim&\frac{-i}{\epsilon'(\epsilon )}\delta_{\Pi^j}+\frac{1}{1+\epsilon'(\epsilon )}A_j
\;\;, \end{eqnarray} 
where relative signs matter, while constant factors have been omitted. 
Keeping $\epsilon'$ small but {\it finite}, we collect the resulting correction 
terms for $\hat{\cal H}$, which accordingly deform the QFT: 
\begin{equation}\label{Heps}
\hat{\cal H}_{\epsilon'}
\equiv \epsilon'e'_0\Big (
i\hat\Pi\gamma^0\gamma^jT\gamma_5C\hat\psi (a'_j-\tilde a'_j) 
-\hat\psi^\dagger\gamma^0\gamma^j\hat\psi\tilde a'_j 
+\hat\Pi^\dagger\gamma^0\gamma^j\hat\Pi a'_j\Big ) 
\;\;. \end{equation}
With $\epsilon'$ spatially homogeneous, the $F\tilde F$-terms 
from Eq.\,(\ref{HemA3}) are absent.  

The second and third terms are of the usual ``$j\cdot A$'' form. However, the visible  
current couples to the ghost vector potential and the ghost current couples to the visible vector potential. 
These lead to {\it vacuum decay}, thereby lowering the total energy of the system indefinitely. 
Similarly as the visible-ghost coupling induced by graviton loops, the topic of  
Ref.\,\cite{KS05}, the situation here could even be phaenomenologically acceptable, 
if the effective coupling $\epsilon'e'_0$ is sufficiently small. -- 
Meanwhile, the first term of $\hat{\cal H}_{\epsilon'}$ couples a transition current of visible 
and ghost charges to the 
visible and ghost vector fields. Like the other two, it violates U(1$)_{vis}$ or U(1$)_{gho}$ separately, but 
leaves the overall local U(1) symmetry intact. Therefore, limits on violation of 
{\it charge conservation} in the visible sector could also constrain the size  
of the effective coupling. -- In any case, the present model should not be applied to  
phaenomenology directly. We are rather interested in its new structural   
features which might be reflected in more realistic theories. 

We have seen in Section\,3.2 that the first term of $\hat{\cal H}_{\epsilon'}$ is {\it not} Hermitian. Concerning the quantum limit, it introduces a 
{\it decoherence mechanism}. This may be wellcome as necessary ingredient 
for attempts to solve the measurement problem and, 
more specifically, the problem of reduction or wave function collapse 
\cite{Adler,BZW}. 
Non-unitarity also appears related to the ``information loss'' deemed 
necessary by 't\,Hooft, in order to 
base quantum theory on deterministic dynamics \cite{tHooft01,Vitiello01,Blasone04,tHooft03}. 
-- Presently, the dissipative interaction arises in the deformation of the quantum 
field theory which relates it to the classical ensemble theory with given symmetries.  

\subsection{Locality or How to circumvent Bell's theorem}
From $\hat{\cal H}\equiv\int\mbox{d}^3x\;\hat{\cal H}(x)$ we can read off 
the Hamiltonian density. Explicit calculation  
then shows: 
\begin{equation}\label{loc} 
[\hat{\cal H}(x),\hat{\cal H}(x')]=0\;\;,\;\;\;\mbox{for}\;x\neq x'
\;\;. \end{equation} 
This establishes the locality of the dynamics described by the generalized Hamiltonian, 
Eqs.\,(\ref{Hfinal})--(\ref{Hint}), and especially by $\hat{\cal H}+\hat{\cal H}_{\epsilon'}$ 
from Eqs.\,(\ref{Hamop}) and (\ref{Heps}).       
-- However, this also raises the question how the quantum theory of Section\,3.3  
could possibly emerge from the generalized classical ensemble 
theory developed before. It appears to contradict Bell's theorem which 
rules out {\it local} hidden variable theories \cite{Adler,BZW}.   
  
Two aspects come into play here. Foremost, our ensemble theory 
{\it is nonlocal} with respect to variables of the underlying model --  
a common feature with other emergent quantum models \cite{tHooft01,I05susy,ES02,Smolin}.  
In particular, in 
Eq.\,(\ref{Fourier}), the canonical momentum $\Pi_i$ is traded for 
the vector potential $A_i'$ via the Fourier transform. We remark that 
even without interactions the ensemble theory for the gauge field turns into the  
corresponding free QFT plus ghost copy. 

For the fermionic fields, where 
nothing like an integral transform has been applied, the anticommutativity of 
the ``pseudoclassical'' \cite{CB,FDeW} Grassmann variables is sufficient. 
We recall that the functional Schr\"odinger equation, Eq.\,(\ref{Schroedinger}), 
follows entirely from the classical Liouville equation, with added 
diffusion term $\hat{\cal H}_{int}$. However, we have found in Section\,3.2       
that the ensemble theory which smoothly connects to quantum theory 
lives in a larger Hilbert space 
than the classical one, concerning the fermionic fields.  

We conclude here that our results do not contradict Bell's theorem. 

\section{Conclusions}
Presently it has been demonstrated    
that a classical ensemble theory 
can be deformed into a QFT in agreement with the 
KS scenario involving energy-parity symmetry \cite{KS05}.   
-- We have related this deformation
to a ``varying alpha'' coupling in an Abelian U(1) gauge theory 
\cite{B82,Uzan03,KM04,Orfeu04}. -- 
Founded on a set of physical axioms, the existence of a one-parameter deformation 
has been shown which connects the nonrelativistic Schr\"odinger theory 
with a classical ensemble theory 
\cite{Parwani}. Here we provide a relativistic field theory realization of such a 
deformation.   

Our approach has incorporated an element of nonlocality, which is essential for the emerging bosonic 
variables. It also employs the ``pseudoclassical mechanics'' based on anticommuting Grassmann 
variables \cite{CB}, in order to include classical charged particles with spin which 
emerge as fermions. Combined with the Hilbert space 
formalism for the phase space description of a {\it classical system} 
\cite{tHooft06}, this has led, 
in a particular limit of the coupling, to the corresponding {\it quantum field theory}.  
We have discussed necessary dissipative and diffusive interactions mediating between classical 
and quantum limits, which imply a decoherence mechanism.   

A number of open problems and interesting topics for future study arise. --  
We have pointed out the larger-than-classical Hilbert space related 
to the realization of fermionic function space and operators. 
Had we considered charged scalar fields instead, 
this feature were absent, and a model like scalar QED plus ghost copy results  
directly from a classical Hermitian ensemble theory. Therefore, it will be interesting  
to further deconstruct fermions in terms of classical concepts within the present approach. --  
Regarding more realistic models, an  
extension to non-Abelian gauge theories seems very important, 
since nonlinear selfinteractions might obstruct energy-parity and symmetry doubling. -- 
Since we assumed a given deformation parameter, this raises the question whether 
a selfconsistent model can be built, dealing with gravity and 
closer in spirit to the KS scenario, which 
partly motivated the present study \cite{KS05}.   

Ending on a speculative note, if a wider range of deterministic quantum models can be 
constructed, this will likely challenge current ideas about space-time at the Planck scale. 
Perhaps energy-parity will play a role in 
this and in solving the old cosmological constant problem.   

\section*{Acknowledgements} 
I thank A.\,DiGiacomo, O.\,Bertolami, D.\,Oriti, F.\,Markopoulou, C.\,Kiefer, 
and G.\,Vitiello for discussions and R.\,Erdem, S.\,Hossenfelder, and 
S.\,Sarkar for correspondence. I thank G.\,'t\,Hooft for making a 
preliminary version of Ref.\,\cite{tHooft06} available to me and for related 
discussions. 



\end{document}